\documentclass[12pt,a4paper]{article}

\usepackage{lineno}  % for line numbering during review
\usepackage{graphicx}  % to include figures (can also use other packages)
\usepackage{xspace} % To avoid problems with missing or double spaces after
\usepackage{color}
\usepackage{colortbl}
\usepackage{amsmath} % Adds a large collection of math symbols
\usepackage{ifthen} % for conditional statements
\graphicspath{{./figs/}} % Make Latex search fig subdir for figures

\newboolean{pdflatex}
\setboolean{pdflatex}{false} % use this if using eps figures
\newboolean{articletitles}
\setboolean{articletitles}{true} % False removes titles in references

\newboolean{uprightparticles}
\setboolean{uprightparticles}{false} %Set to true to get roman particle symbols

\usepackage{amssymb}
\usepackage{amsfonts}
\usepackage{upgreek} % Adds in support for greek letters in roman typeset

\usepackage{hyperref}    % Hyperlinks in references
\usepackage[all]{hypcap} % Internal hyperlinks to floats.

\def\lhcb {LHCb\xspace}
\def\ux85 {UX85\xspace}

\def\atlas {ATLAS\xspace}

\def\cdf    {CDF\xspace}
\def\dzero  {D0\xspace}

\def\herab  {HERA-B\xspace}

\ifthenelse{\boolean{uprightparticles}}%
{
 
 \def\Pgamma      {\ensuremath{\upgamma}\xspace}

 \def\Pmu         {\ensuremath{\upmu}\xspace}

 \def\Pchi        {\ensuremath{\upchi}\xspace}                 
 \def\Ppsi        {\ensuremath{\uppsi}\xspace}

 \def\PDelta      {\ensuremath{\Delta}\xspace}                 
 \def\PXi      {\ensuremath{\Xi}\xspace}                 
 \def\PLambda      {\ensuremath{\Lambda}\xspace}                 
 \def\PSigma      {\ensuremath{\Sigma}\xspace}                 
 \def\POmega      {\ensuremath{\Omega}\xspace}                 
 \def\PUpsilon      {\ensuremath{\Upsilon}\xspace}                 
 
 \def\PB      {\ensuremath{\mathrm{B}}\xspace}                 
                  
 \def\PD      {\ensuremath{\mathrm{D}}\xspace}

 \def\PJ      {\ensuremath{\mathrm{J}}\xspace}                 
 \def\PK      {\ensuremath{\mathrm{K}}\xspace}

 \def\Pb      {\ensuremath{\mathrm{b}}\xspace}                 
 \def\Pc      {\ensuremath{\mathrm{c}}\xspace}

 \def\Pi      {\ensuremath{\mathrm{i}}\xspace}

}
{
 
 \def\Pgamma      {\ensuremath{\gamma}\xspace}

 \def\Pmu         {\ensuremath{\mu}\xspace}

 \def\Pchi        {\ensuremath{\chi}\xspace}                 
 \def\Ppsi        {\ensuremath{\psi}\xspace}                 
                  
 \mathchardef\PDelta="7101
 \mathchardef\PXi="7104
 \mathchardef\PLambda="7103
 \mathchardef\PSigma="7106
 \mathchardef\POmega="710A
 \mathchardef\PUpsilon="7107
                  
 \def\PB      {\ensuremath{B}\xspace}                 
                  
 \def\PD      {\ensuremath{D}\xspace}

 \def\PJ      {\ensuremath{J}\xspace}                 
 \def\PK      {\ensuremath{K}\xspace}

 \def\Pb      {\ensuremath{b}\xspace}                 
 \def\Pc      {\ensuremath{c}\xspace}

 \def\Pi      {\ensuremath{i}\xspace}

}

\def\mumu       {\ensuremath{\Pmu^+\Pmu^-}\xspace}

\def\g      {\ensuremath{\Pgamma}\xspace}

\def\cquark    {\ensuremath{\Pc}\xspace}

\def\bquark    {\ensuremath{\Pb}\xspace}

\def\kaon  {\ensuremath{\PK}\xspace}
  \def\Kbar  {\kern 0.2em\overline{\kern -0.2em \PK}{}\xspace}

\def\Kz    {\ensuremath{\kaon^0}\xspace}
\def\Kzb   {\ensuremath{\Kbar^0}\xspace}
\def\KzKzb {\ensuremath{\Kz \kern -0.16em \Kzb}\xspace}
\def\Kp    {\ensuremath{\kaon^+}\xspace}
\def\Km    {\ensuremath{\kaon^-}\xspace}

\def\KpKm  {\ensuremath{\Kp \kern -0.16em \Km}\xspace}

  \def\Dbar    {\kern 0.2em\overline{\kern -0.2em \PD}{}\xspace}
\def\D       {\ensuremath{\PD}\xspace}

\def\Dz      {\ensuremath{\D^0}\xspace}
\def\Dzb     {\ensuremath{\Dbar^0}\xspace}
\def\DzDzb   {\ensuremath{\Dz {\kern -0.16em \Dzb}}\xspace}
\def\Dp      {\ensuremath{\D^+}\xspace}
\def\Dm      {\ensuremath{\D^-}\xspace}

\def\DpDm    {\ensuremath{\Dp {\kern -0.16em \Dm}}\xspace}

  \def\Bbar    {\kern 0.18em\overline{\kern -0.18em \PB}{}\xspace}

\def\jpsi     {\ensuremath{{\PJ\mskip -3mu/\mskip -2mu\Ppsi\mskip 2mu}}\xspace}
\def\psitwos  {\ensuremath{\Ppsi{(2S)}}\xspace}

  \def\Y#1S{\ensuremath{\PUpsilon{(#1S)}}\xspace}% no space before {...}!

\def\chib#1P{\ensuremath{\Pchi_{\bquark}{(#1P)}}\xspace}
\def\chibJ#1P{\ensuremath{\Pchi_{\bquark{}J}{(#1P)}}\xspace}

\def\Lbar {\ensuremath{\kern 0.1em\overline{\kern -0.1em\Lambda\kern -0.05em}\kern 0.05em{}}\xspace}

\def\to                 {\ensuremath{\rightarrow}\xspace}

\def\AT#1     {\ensuremath{A_{\mathrm{T}}^{#1}}\xspace}           % 2

\def\C#1      {\ensuremath{\mathcal{C}_{#1}}\xspace}                       % 9
\def\Cp#1     {\ensuremath{\mathcal{C}_{#1}^{'}}\xspace}                    % 7
\def\Ceff#1   {\ensuremath{\mathcal{C}_{#1}^{\mathrm{(eff)}}}\xspace}        % 9  
\def\Cpeff#1  {\ensuremath{\mathcal{C}_{#1}^{'\mathrm{(eff)}}}\xspace}       % 7
\def\Ope#1    {\ensuremath{\mathcal{O}_{#1}}\xspace}                       % 2
\def\Opep#1   {\ensuremath{\mathcal{O}_{#1}^{'}}\xspace}                    % 7

\newcommand{\tev}{\ensuremath{\mathrm{\,Te\kern -0.1em V}}\xspace}
\newcommand{\gev}{\ensuremath{\mathrm{\,Ge\kern -0.1em V}}\xspace}
\newcommand{\mev}{\ensuremath{\mathrm{\,Me\kern -0.1em V}}\xspace}
\newcommand{\kev}{\ensuremath{\mathrm{\,ke\kern -0.1em V}}\xspace}
\newcommand{\ev}{\ensuremath{\mathrm{\,e\kern -0.1em V}}\xspace}
\newcommand{\gevc}{\ensuremath{{\mathrm{\,Ge\kern -0.1em V\!/}c}}\xspace}
\newcommand{\mevc}{\ensuremath{{\mathrm{\,Me\kern -0.1em V\!/}c}}\xspace}
\newcommand{\gevcc}{\ensuremath{{\mathrm{\,Ge\kern -0.1em V\!/}c^2}}\xspace}
\newcommand{\gevgevcccc}{\ensuremath{{\mathrm{\,Ge\kern -0.1em V^2\!/}c^4}}\xspace}
\newcommand{\mevcc}{\ensuremath{{\mathrm{\,Me\kern -0.1em V\!/}c^2}}\xspace}

\def\mm   {\ensuremath{\rm \,mm}\xspace}

\def\mum  {\ensuremath{\,\upmu\rm m}\xspace}

\def\invpb {\ensuremath{\mbox{\,pb}^{-1}}\xspace}

\def\gsim{{~\raise.15em\hbox{$>$}\kern-.85em
          \lower.35em\hbox{$\sim$}~}\xspace}
\def\lsim{{~\raise.15em\hbox{$<$}\kern-.85em
          \lower.35em\hbox{$\sim$}~}\xspace}

\def\pt         {\mbox{$p_{\rm T}$}\xspace}

\def\DM         {\ensuremath{\Delta M}}

\def\evtgen     {\mbox{\textsc{EvtGen}}\xspace}
\def\pythia     {\mbox{\textsc{Pythia}}\xspace}

\def\geant      {\mbox{\textsc{Geant4}}\xspace}

\def\photos     {\mbox{\textsc{Photos}}\xspace}

\def\tell1  {TELL1\xspace}
\def\ukl1   {UKL1\xspace}

\usepackage{cite} % Allows for ranges in citations
\usepackage{mciteplus}

\begin{document}

%%%%%%%%%%%%%%%%%%%%%%%%%
%%%%% Title     %%%%%%%%%
%%%%%%%%%%%%%%%%%%%%%%%%%
\renewcommand{\thefootnote}{\fnsymbol{footnote}}
\setcounter{footnote}{1}

% %%%%%%% CHOOSE --------
% \input{title-LHCb-ANA}
% \input{title-LHCb-CONF}
% $Id: title-LHCb-PAPER.tex 16065 2012-02-20 22:49:03Z uegede $
% ===============================================================================
% Purpose: LHCb-PAPER journal paper title page template
% Author: 
% Created on: 2010-09-25
% ===============================================================================

%%%%%%%%%%%%%%%%%%%%%%%%%
%%%%%  TITLE PAGE  %%%%%%
%%%%%%%%%%%%%%%%%%%%%%%%%
\begin{titlepage}
\pagenumbering{roman}

% Header ---------------------------------------------------
\vspace*{-1.5cm}
\centerline{\large EUROPEAN ORGANIZATION FOR NUCLEAR RESEARCH (CERN)}
\vspace*{1.5cm}
\hspace*{-0.5cm}
\begin{tabular*}{\linewidth}{lc@{\extracolsep{\fill}}r}
\ifthenelse{\boolean{pdflatex}}% Logo format choice
{\vspace*{-2.7cm}\mbox{\!\!\!\includegraphics[width=.14\textwidth]{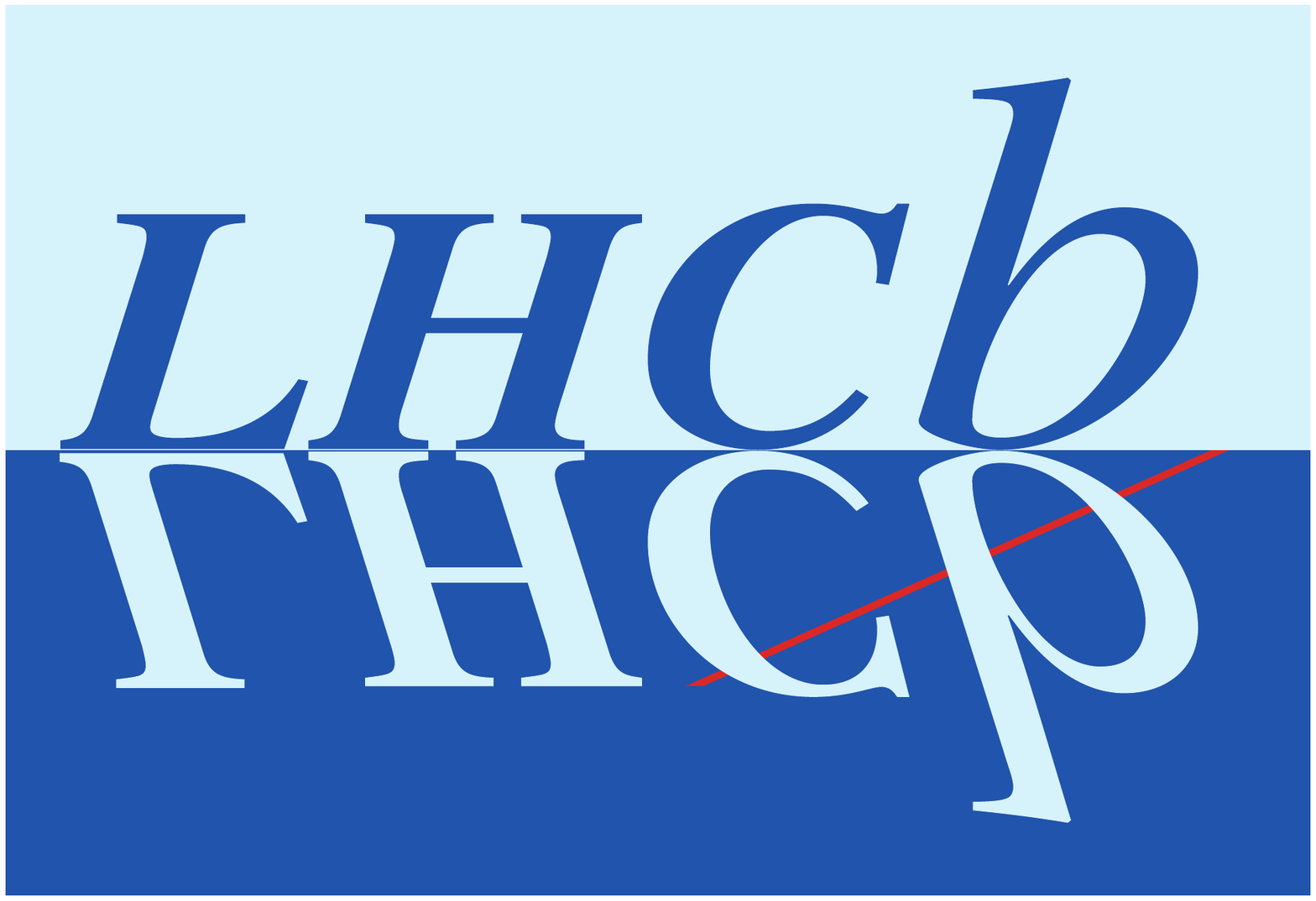}} & &}%
{\vspace*{-1.2cm}\mbox{\!\!\!\includegraphics[width=.12\textwidth]{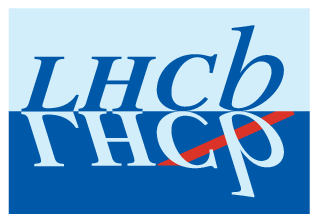}} & &}%
\\
 & & CERN-PH-EP-2012-237 \\  % ID 
 & & LHCb-PAPER-2012-015 \\  % ID 
 & & \today \\ % Date - Can also hardwire e.g.: 23 March 2010
% & & version 2.6\\
% not in paper \hline
\end{tabular*}

\vspace*{4.0cm}

% Title --------------------------------------------------
{\bf\boldmath\huge
\begin{center}
Measurement of the fraction of \Y1S originating from \chib1P decays
in $pp$ collisions at $\sqrt{s} = 7\tev$
\end{center}
}

\vspace*{2.0cm}

% Authors -------------------------------------------------
\begin{center}
The \lhcb collaboration\footnote{Authors are listed on the following pages.}
\end{center}

\vspace{\fill}

% Abstract -----------------------------------------------
\begin{abstract}
  \noindent
The production of \chib1P mesons in $pp$ collisions at a centre-of-mass
energy of $7\tev$ is studied using $32\invpb$ of data collected with the
\lhcb detector. The $\chib1P$ mesons are reconstructed in the decay mode
$\chib1P \to \Y1S\g \to \mumu\g$. The fraction of \Y1S originating from
\chib1P decays in the \Y1S transverse momentum range
$6 < \pt^{\Y1S} < 15\gevc$ and rapidity range $2.0 < y^{\Y1S} < 4.5$ is
measured to be $(20.7\pm 5.7\pm 2.1^{+2.7}_{-5.4})\%$, where the first
uncertainty is statistical, the second is systematic and the last gives
the range of the result due to the unknown \Y1S and \chib1P
polarizations.
\end{abstract}

\vspace*{2.0cm}

\begin{center}
Submitted to JHEP
\end{center}

\vspace{\fill}

\end{titlepage}

%%%%%%%%%%%%%%%%%%%%%%%%%%%%%%%%
%%%%%  EOD OF TITLE PAGE  %%%%%%
%%%%%%%%%%%%%%%%%%%%%%%%%%%%%%%%

%  empty page follows the title page ----
\newpage
\setcounter{page}{2}
\mbox{~}
\newpage

% Author List ----------------------------
%  You need to get a new author list!
%%%%%%%%%%%%%%%%%%%%%%%%%%%%%%%%%%%%%%%%%%
\centerline{\large\bf LHCb collaboration}
\begin{flushleft}
\small
R.~Aaij$^{38}$, 
C.~Abellan~Beteta$^{33,n}$, 
A.~Adametz$^{11}$, 
B.~Adeva$^{34}$, 
M.~Adinolfi$^{43}$, 
C.~Adrover$^{6}$, 
A.~Affolder$^{49}$, 
Z.~Ajaltouni$^{5}$, 
J.~Albrecht$^{35}$, 
F.~Alessio$^{35}$
% ...

{\footnotesize \it
% ...
$ ^{5}$Clermont Universit\'{e}, Universit\'{e} Blaise Pascal, CNRS/IN2P3, LPC, Clermont-Ferrand, France\\
$ ^{6}$CPPM, Aix-Marseille Universit\'{e}, CNRS/IN2P3, Marseille, France\\
% ...
$ ^{11}$Physikalisches Institut, Ruprecht-Karls-Universit\"{a}t Heidelberg, Heidelberg, Germany\\
% ...
$ ^{33}$Universitat de Barcelona, Barcelona, Spain\\
$ ^{34}$Universidad de Santiago de Compostela, Santiago de Compostela, Spain\\
$ ^{35}$European Organization for Nuclear Research (CERN), Geneva, Switzerland\\
% ...
$ ^{38}$Nikhef National Institute for Subatomic Physics, Amsterdam, The Netherlands\\
% ...
$ ^{43}$H.H. Wills Physics Laboratory, University of Bristol, Bristol, United Kingdom\\
% ...
$ ^{49}$Oliver Lodge Laboratory, University of Liverpool, Liverpool, United Kingdom\\
% ...
$ ^{n}$LIFAELS, La Salle, Universitat Ramon Llull, Barcelona, Spain\\
% ...
}
\end{flushleft}
%%%%%%%%%%%%%%%%%%%%%%%%%%%%%%%%%%%%%%%%%%

\cleardoublepage

% %%%%%%%%%%%%% ---------

\renewcommand{\thefootnote}{\arabic{footnote}}
\setcounter{footnote}{0}

%%%%%%%%%%%%%%%%%%%%%%%%%%%%%%%%
%%%%%  Table of Content   %%%%%%
%%%%%%%%%%%%%%%%%%%%%%%%%%%%%%%%
%%%% Uncomment next 2 lines if desired
%\tableofcontents
%\cleardoublepage

%%%%%%%%%%%%%%%%%%%%%%%%%
%%%%% Main text %%%%%%%%%
%%%%%%%%%%%%%%%%%%%%%%%%%

\pagestyle{plain} % restore page numbers for the main text
\setcounter{page}{1}
\pagenumbering{arabic}

% %%%%%%% CHOOSE --------
%% ----------------------------------
%% Line numbering on the left margin 
%% ----------------------------------
%% Uncomment during review phase. 
%% Comment it out before a final submission.
%\linenumbers
%% --------------------------------
% %%%%%%%%%%%%% ---------

% You can include short sections directly in the main tex file.
% However, for larger papers it is desirable to split the text into
% several semiautonomous files, which can be revised independently.
% This is especially useful when developing a document in
% collaboration with several people, since then different parts can be
% edited independently.  This type of file organization is shown here.
% 

\section{Introduction}
\label{sec:Introduction}
The production of heavy quarkonium states at hadron colliders is a subject
of experimental and theoretical interest~\cite{Brambilla:2010cs}. The
non-relativistic QCD (NRQCD) factorization approach has been developed to
describe the inclusive production and decay of quarkonia~\cite{Bodwin}.
The \lhcb experiment has measured the production of inclusive
$\jpsi\to\mumu$~\cite{LHCb-PAPER-2011-003},
$\psitwos$~\cite{LHCb-PAPER-2011-045} and $\PUpsilon(nS)\to\mumu$
$(n = 1,2,3)$~\cite{LHCb-PAPER-2011-036} mesons in $pp$ collisions as
a function of the quarkonium transverse momentum \pt and rapidity $y$ over
the range $0 < \pt < 15\gevc$ and $2.0 < y < 4.5$. A significant fraction
of the cross-section for both \jpsi and $\PUpsilon(nS)$ production is expected
to be due to feed-down from higher quarkonium states. Understanding the
size of this effect is important for the interpretation of the quarkonia
cross-section and polarization data. A few experimental studies
of hadroproduction of $P$-wave quarkonia have been reported. In the case
of the $\chi_{cJ}$ states, with spin $J=0,1,2$, measurements from the
\cdf~\cite{CDF-chic,CDF-Jpsifromchic}, \herab~\cite{HERAB-chic} and
\lhcb~\cite{LHCb-PAPER-2011-019,LHCb-PAPER-2011-030} experiments exist,
while $\chi_{bJ}$ related measurements have been reported by the
\cdf~\cite{CDF-chib}, \atlas~\cite{ATLAS-chib3P} and
\dzero~\cite{d0-chib} experiments.

This paper reports studies of the inclusive production of the $P$-wave
\chibJ1P states, collectively referred to as \chib1P throughout the paper.
The \chib1P mesons are reconstructed through the radiative decay
$\chib1P\to\Y1S\g$ in the \Y1S rapidity and transverse momentum range
$2.0 < y^{\Y1S} < 4.5$ and $6 < \pt^{\Y1S} < 15\gevc$. The $\chi_{b2}$ and
$\chi_{b1}$ states differ in mass by $20\mevcc$ and the $\chi_{b1}$ and
$\chi_{b0}$ states by $33\mevcc$~\cite{Nakamura:2010zzi}. Since these
differences are comparable with the experimental resolution, the total
fraction of \Y1S originating from \chib1P decays is reported. The results
presented here use a data sample collected at the LHC with the \lhcb
detector at a centre-of-mass energy of $7\tev$ and correspond to an
integrated luminosity of $32\invpb$.

\section{\lhcb detector}
\label{sec:detector}
The \lhcb detector~\cite{Alves:2008zz} is a single-arm forward
spectrometer covering the pseudorapidity range $2<\eta <5$, designed
for the study of particles containing \bquark or \cquark quarks. The
detector includes a high precision tracking system consisting of a
silicon-strip vertex detector surrounding the $pp$ interaction region,
a large-area silicon-strip detector located upstream of a dipole
magnet with a bending power of about $4{\rm\,Tm}$, and three stations
of silicon-strip detectors and straw drift tubes placed
downstream. The combined tracking system has a momentum resolution
$\Delta p/p$ that varies from 0.4\% at 5\gevc to 0.6\% at 100\gevc,
and an impact parameter resolution of 20\mum for tracks with high
transverse momentum (\pt). Charged hadrons are identified using two
ring-imaging Cherenkov detectors. Photon, electron and hadron
candidates are identified by a calorimeter system consisting of
scintillating-pad and preshower detectors, an electromagnetic calorimeter
and a hadronic calorimeter. Muons are identified by a system composed of
alternating layers of iron and multiwire proportional chambers.
The nominal detector performance for photons and muons is described
in~\cite{Alves:2008zz}. The processes of radiative transitions of
$\chi_{cJ} \to J/\psi\gamma$, $J=1,2$ with similar kinematics of the
photons are studied in~\cite{LHCb-PAPER-2011-019,LHCb-PAPER-2011-030}.
Another physical analysis which uses $\pi^0\to\gamma\gamma$,
$\eta\to\gamma\gamma$ and $\eta'\to\rho^0\gamma$ is available
as~\cite{LHCb-PAPER-2012-022}.

The trigger consists of a hardware stage followed by a software stage
which applies a full event reconstruction. The trigger used for this
analysis selects a pair of oppositely-charged muon candidates, where
either one of the muons has a $\pt > 1.8\gevc$ or one of the pair has a
$\pt > 0.56\gevc$ and the other has a $\pt > 0.48\gevc$. The invariant
mass of the pair is required to be greater than $2.9\gevcc$. The photons
are not used in the trigger decision.

For the simulation, $pp$ collisions are generated using
\pythia~6.4~\cite{Sjostrand:2006za} with a specific \lhcb
configuration~\cite{LHCb-PROC-2010-056}.  Decays of hadronic particles
are described by \evtgen~\cite{Lange:2001uf} in which final state
radiation is generated using \photos~\cite{Golonka:2005pn}. The
interaction of the generated particles with the detector and its
response are implemented using the \geant
toolkit~\cite{Allison:2006ve, *Agostinelli:2002hh} as described in
Ref.~\cite{LHCb-PROC-2011-006}. The simulated signal events contain at
least one $\Y1S\to\mumu$ decay with both muons in the \lhcb acceptance.
In this sample of simulated events the fraction of \Y1S mesons produced
in \chib1P decays is 47\% and both the \chib1P and \Y1S mesons are
produced unpolarized.

\section{Event selection}
\label{sec:selection}
The reconstruction of the \chib1P meson proceeds via the identification
of an \Y1S meson combined with a reconstructed photon. The $\PUpsilon(nS)$
candidates are formed from a pair of oppositely-charged tracks that are
identified as muons. Each track is required to have a good track fit
quality. The two muons are required to originate from a common vertex with
a distance to the primary vertex less than $1\mm$.

\begin{figure}[h]
 \begin{center}
  \ifthenelse{\boolean{pdflatex}}
  {\includegraphics*[width=0.7\textwidth]{figs/m2mu.pdf}}
  {\includegraphics*[width=0.7\textwidth]{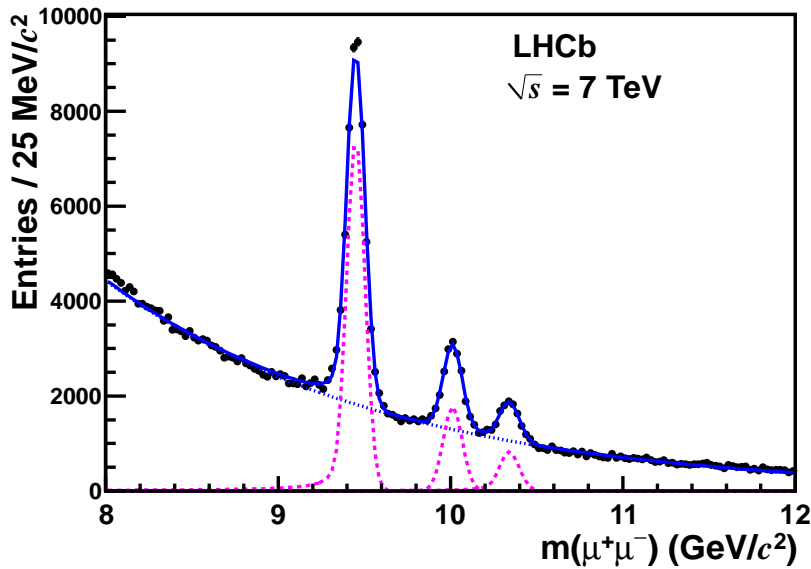}}
 \end{center}
 \caption{\label{fg:m2mu}\small Distribution of the $\mumu$ mass for
selected $\PUpsilon(nS)$ candidates (black points), together with the
result of the fit (solid blue curve), including the background (dotted blue
curve) and the signal (dashed magenta curve) contributions.}
\end{figure}

The invariant mass distribution of the $\mumu$ candidates is shown in
Fig.~\ref{fg:m2mu}. It is modelled with the sum of three Crystal Ball
functions~\cite{Skwarnicki:1986xj}, describing the \Y1S, \Y2S and \Y3S
signals, and an exponential function for the combinatorial background. The
parameters of the Crystal Ball functions that describe the radiative tail
of the \Y1S, \Y2S and \Y3S mass distributions are fixed to the values $a=2$
and $n=1$~\cite{LHCb-PAPER-2011-036}. The measured \Y1S signal yield, mass
and width are $N_{\Y1S}=39\,635\pm 252$, $m_{\Y1S}=9449.2\pm 0.4\mevcc$ and
$\sigma_{\Y1S}=51.7\pm 0.4\mevcc$, where the uncertainties are statistical
only.

%\section{$\chi_b \to \Y1S\g$ selection}
%\label{sec:chib}
The \Y1S candidates with a $\pt^{\Y1S} > 6\gevc$ and a $\mumu$ invariant
mass in the range $9.36 - 9.56\gevcc$ are combined with photons to form
\chib1P candidates. The photons are required to have
$\pt^\gamma > 0.6\gevc$ and $\cos\theta^*_\g > 0$, where $\theta^*_\g$ is
the angle of the photon direction in the centre-of-mass frame of the
$\mumu\g$ system with respect to the momentum of this system in the
laboratory frame.

\begin{figure}[htbp]
 \begin{center}
  \ifthenelse{\boolean{pdflatex}}
  {\includegraphics*[width=0.7\textwidth]{figs/dm2mug.pdf}}
  {\includegraphics*[width=0.7\textwidth]{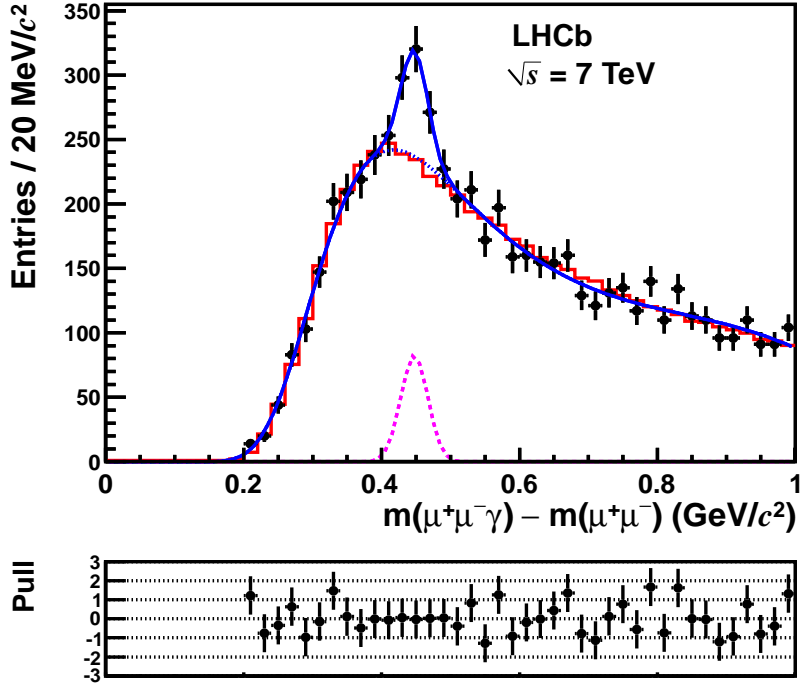}}
 \end{center}
 \caption{\label{fg:dm2mug}\small Distribution of the mass difference
$m(\mumu\g) - m(\mumu)$ for selected \chib1P candidates (black points),
together with the result of the fit (solid blue curve), including
background (dotted blue curve) and signal (dashed magenta curve)
contributions. The solid (red) histogram is an alternative background
estimation using simulated events containing a \Y1S that does not
originate from a \chib1P decay, normalized to the data. It is used for
evaluation of the systematic uncertainty due to the choice of fitting
model. The bottom insert shows the pull distribution of the fit.
The pull is defined as the difference between the data and fit value
divided by the data error.}
\end{figure}

The $\chib1P$ signal peak observed in the distribution of the mass
difference, $x = m(\mumu\g) - m(\mumu)$, is shown in Fig.~\ref{fg:dm2mug}
for the range $6 < \pt^{\Y1S} < 15\gevc$. It is modelled with an empirical
function given by

\begin{equation}
\frac{dN}{dx} = A_1\frac{1}{\sqrt{2\pi}\sigma}e^{-\frac{(x-\DM)^2}{2\sigma^2}} + A_2(x-x_0)^\alpha e^{-(c_1x +c_2x^2 +c_3x^3)},\label{eq:fitfun}
\end{equation}
where $A_1$, $\DM$, $\sigma$, $A_2$, $x_0$, $\alpha$, $c_1$, $c_2$ and
$c_3$ are free parameters. The Gaussian function describes the signal and
the second term models the background. The number of \chib1P signal decays
obtained from the fit is $201\pm 55$. The mean value of the Gaussian
function is $447\pm 4\mevcc$ and its width is $19.0\pm 4.2\mevcc$. The
expected values of the mass differences for the three \chibJ1P states are
$\DM(\chi_{b2})=452\mevcc$, $\DM(\chi_{b1})=432\mevcc$ and
$\DM(\chi_{b0})=399\mevcc$~\cite{Nakamura:2010zzi}. The peak position in
the data lies between $\DM(\chi_{b2})$ and $\DM(\chi_{b0})$ as expected
for any mixture of $\chibJ1P$ states.

\section{\boldmath Fraction of \Y1S originating from \chib1P decays}
\label{sec:fraction}
The fraction of \Y1S originating from \chib1P decays is determined using
the following assumptions. Firstly, all \Y1S originating from \chib1P
arise from the radiative decay $\chib1P\to\Y1S\g$. Secondly, the total
efficiency for $\Y1S\to\mumu$ as a function of $\pt^{\Y1S}$ is the same
for directly produced \Y1S and for those from feed-down from \chib1P. The
total efficiency includes trigger, detection, reconstruction and
selection. Thirdly, the photon detection, reconstruction and selection
are independent of the $\Y1S\to\mumu$. Hence the total efficiency for
\chib1P is factorized as
$\epsilon_\mathrm{tot}(\chi_b) = \epsilon_\mathrm{cond}(\chi_b)
\cdot \epsilon_\mathrm{tot}(\PUpsilon)$, where
$\epsilon_\mathrm{tot}(\PUpsilon)$ is the total efficiency for \Y1S
and $\epsilon_\mathrm{cond}(\chi_b)$
is the conditional efficiency for \chib1P reconstruction and selection
after the $\Y1S\to\mumu$ candidate has been selected.

The second assumption is tested by comparing the \Y1S efficiencies
obtained using simulated events for direct \Y1S and for \Y1S coming from
decays of \chib1P states. These efficiencies for each $\pt^{\Y1S}$
interval agree within the statistical error, which is less than 0.5\%.

The conditional \chib1P reconstruction and selection efficiency is
estimated from simulation as

\def\NMCrecchib{\ensuremath{N^\mathrm{MC}_\mathrm{rec}(\Pchi_{\bquark})}\xspace}
\def\NMCrecY{\ensuremath{N^\mathrm{MC}_\mathrm{rec}(\PUpsilon)}\xspace}
\def\NMCgenchib{\ensuremath{N^\mathrm{MC}_\mathrm{gen}(\Pchi_{\bquark})}\xspace}
\def\NMCgenY{\ensuremath{N^\mathrm{MC}_\mathrm{gen}(\PUpsilon)}\xspace}
\begin{equation}
\epsilon_\mathrm{cond}(\chi_b) =
\frac{\epsilon_\mathrm{tot}(\chi_b)}{\epsilon_\mathrm{tot}(\PUpsilon)} =
\frac{\NMCrecchib}{\NMCgenchib} \cdot \frac{\NMCgenY}{\NMCrecY},
\end{equation}
where \NMCrecchib and \NMCrecY are the number of \chib1P and \Y1S mesons
obtained from the fit, and \NMCgenchib and \NMCgenY are the number of
generated \chib1P and \Y1S mesons, respectively. The value obtained is
$\epsilon_\mathrm{cond}(\chi_b) = (9.4\pm 0.1)\%$ for
$6 < \pt^{\Y1S} < 15\gevc$ and $2.0 < y^{\Y1S} < 4.5$.

The fraction of \Y1S originating from \chib1P decays is determined from
the ratio

\def\Nrecchib{\ensuremath{N_\mathrm{rec}(\Pchi_{\bquark})}\xspace}
\def\NrecY{\ensuremath{N_\mathrm{rec}(\PUpsilon)}\xspace}
\def\Ntotchib{\ensuremath{N_\mathrm{prod}(\Pchi_{\bquark})}\xspace}
\def\NtotY{\ensuremath{N_\mathrm{prod}(\PUpsilon)}\xspace}
\begin{equation}
\frac{\Ntotchib}{\NtotY} =
\frac{\Nrecchib/\epsilon_\mathrm{tot}(\chi_b)}
{\NrecY/\epsilon_\mathrm{tot}(\Upsilon)} =
\frac{\Nrecchib/\epsilon_\mathrm{cond}(\chi_b)}{\NrecY},
\end{equation}
where \Ntotchib and \NtotY are the total numbers of $\chib1P\to\Y1S\g$
and \Y1S mesons produced, and \Nrecchib and \NrecY are the numbers of
reconstructed \chib1P and \Y1S mesons obtained from the fits to the data,
respectively. As the muons from the \Y1S are explicitly required to
trigger the event, the efficiency of the trigger cancels in this ratio.
The fraction of \Y1S originating from \chib1P decays for
$6 < \pt^{\Y1S} < 15\gevc$ and $2.0 < y^{\Y1S} < 4.5$ is found to be
$(20.7\pm 5.7)\%$, where the uncertainty is statistical only.

\begin{figure}[htbp]
 \begin{center}
  \ifthenelse{\boolean{pdflatex}}
  {\includegraphics*[width=0.7\textwidth]{figs/fra.pdf}}
  {\includegraphics*[width=0.7\textwidth]{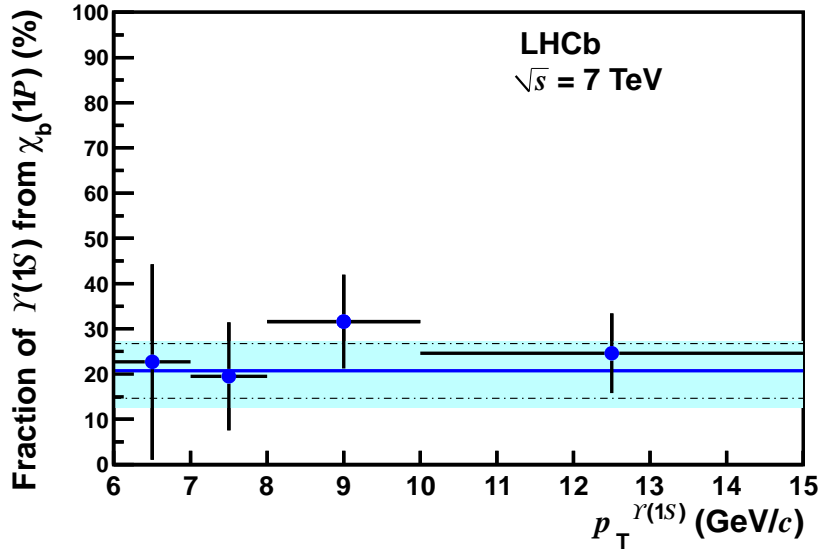}}
 \end{center}
 \caption{\label{fg:fra}\small Fraction of $\Y1S$ originating from
$\chib1P$ decays for different $\pt^{\Y1S}$ bins, assuming production of
unpolarized $\Y1S$ and $\chib1P$ mesons, shown with solid circles. The
vertical error bars are statistical only. The result determined for the
range $6 < \pt < 15\gevc$ is shown with the horizontal solid line, its
statistical uncertainty with the dash-dotted lines, and its total
uncertainty (statistical and systematic, including that due to the unknown
polarization) with the shaded (light blue) band.}
\end{figure}

The procedure is repeated in four bins of $\pt^{\Y1S}$, giving the results
shown in Table~\ref{tb:frac} and Fig.~\ref{fg:fra}. No significant
$\pt^{\Y1S}$ dependence is observed. The mean of the measurements
performed in the individual bins is consistent with the measurement
obtained in the whole $\pt^{\Y1S}$ range.

\begin{table}[t]
\caption{\label{tb:frac}\small Number of reconstructed \chib1P and \Y1S
signal candidates, conditional efficiency and fraction of \Y1S
originating from \chib1P decays for different $\pt^{\Y1S}$ bins. The
uncertainties are statistical only.}
{\small\begin{center}
\noindent\begin{tabular}{lr@{$\,\pm\,$}lr@{$\,\pm\,$}lr@{$\,\pm\,$}lr@{$\,\pm\,$}l|r@{$\,\pm\,$}l}
$\pt^{\Y1S} (\gevc)$                   &\multicolumn{2}{c}{$6-7$}&\multicolumn{2}{c}{$7-8$}&\multicolumn{2}{c}{$8-10$}&\multicolumn{2}{c|}{$10-15$}&\multicolumn{2}{c}{$6-15$}\\ \hline
\Nrecchib                              &$41$  &$39$              &$35$  &$22$              &$91$  &$30$               &$82$  &$29$                 &$201$    &$55$\\
\NrecY                                 &$2730$&$64$              &$2193$&$57$              &$2866$&$64$               &$2627$&$59$                 &$10\,345$&$123$\\
$\epsilon_\mathrm{cond}(\chi_b)$ in \% &$6.7$ &$0.2$             &$8.3$ &$0.2$             &$10.0$&$0.2$              &$12.8$&$0.2$                &$9.4$    &$0.1$\\
Fraction in \%                         &$23$  &$22$              &$20$  &$12$              &$32$  &$10$               &$25$  &$9$                  &$21$     &$6$\\
\end{tabular}
\end{center}}
\end{table}

\section{Systematic uncertainties}
\label{sec:systematic}
Studies of quarkonium decays to two
muons~\cite{LHCb-PAPER-2011-003,LHCb-PAPER-2011-036,LHCb-PAPER-2011-019,LHCb-PAPER-2011-030,LHCb-PAPER-2011-045}
show that the total efficiency depends on the polarization of the vector
meson. The effect of the polarization has been studied by repeating the
estimation of the efficiencies $\epsilon_\mathrm{tot}(\chi_b)$ and
$\epsilon_\mathrm{tot}(\PUpsilon)$ for the extreme \chib1P and \Y1S
polarization scenarios and taking the difference in
$\epsilon_\mathrm{cond}(\chi_b)$ as the systematic uncertainty. The
largest variation is found for the cases of 100\% transverse and
longitudinal polarization of the \Y1S. We assign this relative variation
of $^{+13}_{-26}$\% as the range due to the unknown polarizations.

The systematic effect due to the unknown $\chibJ1P$, $J=0,1,2$ relative
contributions is estimated by varying these fractions in the simulation
in such a way that the peak position of the mixture is equal to the peak
position observed in the data plus or minus its statistical uncertainty.
The maximal relative variation of the result is found to be 7\%. This
value is taken as a systematic uncertainty due to the unknown \chibJ1P
mixture.

The systematic uncertainty due to the photon reconstruction efficiency
is determined by comparing the relative yields of the reconstructed
$B^+\to\jpsi (K^{*+}\to K^+\pi^0)$ and $B^+\to\jpsi K^+$ decays in data
and simulated events. It is assumed that the reconstruction efficiencies
of the two photons from the $\pi^0$ are uncorrelated. The uncertainty on
the photon reconstruction efficiency is studied as a function of $\pt^\g$.
The largest systematic uncertainty is found to be 6\% for photons in the
range $0.6 < \pt^\g < 0.7\gevc$, and is dominated by the uncertainties
of the $B^+$ branching fractions.

The systematic uncertainty due to the choice of the background fit model
is estimated from simulated events containing an \Y1S that does not
originate from the decay of a \chib1P. The distribution of the mass
difference obtained with these events, using the same reconstruction and
selection as for data, is shown in Fig.~\ref{fg:dm2mug}, normalized to
the data below $0.38\gevcc$. It describes rather well the background
contribution above $0.38\gevcc$, both in shape and level. The difference
between the number of data events and the normalized number of simulated
background events in the range $0.38-0.50\gevcc$ gives an estimate of the
signal yield. For $6 < \pt^{\Y1S} <15\gevc$ the signal yield obtained
using this method is $211$ to be compared with $201\pm55$ obtained from
the fit. The procedure is repeated in each $\pt^{\Y1S}$ bin. We also
study the variation of signal yield by changing the normalization range
to $0.0-0.3\gevcc$ or $0.7-1.0\gevcc$. The maximal relative difference
of 5\% is taken as the uncertainty due to the choice of the signal and
background description. Systematic uncertainties are summarized in
Table~\ref{tb:syst}.

\begin{table}[t]
\caption{\label{tb:syst}\small Relative systematic uncertainties on the
fraction of \Y1S originating from \chib1P decays.}
\begin{center}
\begin{tabular}{lr}
Source                                  & Uncertainty (\%) \\ \hline
Unknown \chibJ1P mixture                &  7 \\
Photon reconstruction efficiency        &  6 \\
Signal and background description       &  5 \\ \hline
Quadratic sum of the above              & 10 \\
\end{tabular}
\end{center}
\end{table}

\section{Results and conclusions}
\label{sec:Summary}
The production of \chib1P mesons is observed using data corresponding to
an integrated luminosity of $32\invpb$ collected with the \lhcb detector
in $pp$ collisions at $\sqrt{s} = 7\tev$. The fraction of $\Y1S$
originating from $\chib1P$ decays in the kinematic range
$6 < \pt^{\Y1S} < 15\gevc$ and $2.0 < y^{\Y1S} < 4.5$ is measured to be

%\begin{equation}
$$
(20.7\pm 5.7\pm 2.1^{+2.7}_{-5.4})\%,
$$
%\end{equation}
where the first uncertainty is statistical, the second is systematic and
the last gives the range of the result due to the unknown polarization of
\Y1S and \chib1P mesons.

This result can be compared with the \cdf measurement of
$(27.1\pm 6.9\pm 4.4)\%$~\cite{CDF-chib}, obtained in $p\bar{p}$
collisions at $\sqrt{s}=1.8\tev$ in the kinematic range
$\pt^{\Y1S} > 8\gevc$ and $|\eta^{\Y1S}| < 0.7$.

The \chib1P decays are observed to be a significant source of \Y1S mesons
in $pp$ collisions.  This will need to be taken into account in the
interpretation of the measured \Y1S production cross-section and
polarization.

% Do not include this in analysis note and conference reports
\section*{Acknowledgements}

\noindent We express our gratitude to our colleagues in the CERN accelerator
departments for the excellent performance of the LHC. We thank the
technical and administrative staff at CERN and at the LHCb institutes,
and acknowledge support from the National Agencies: CAPES, CNPq,
FAPERJ and FINEP (Brazil); CERN; NSFC (China); CNRS/IN2P3 (France);
BMBF, DFG, HGF and MPG (Germany); SFI (Ireland); INFN (Italy); FOM and
NWO (The Netherlands); SCSR (Poland); ANCS (Romania); MinES of Russia and
Rosatom (Russia); MICINN, XuntaGal and GENCAT (Spain); SNSF and SER
(Switzerland); NAS Ukraine (Ukraine); STFC (United Kingdom); NSF
(USA). We also acknowledge the support received from the ERC under FP7
and the Region Auvergne.

\addcontentsline{toc}{section}{References}
\bibliographystyle{LHCb}
\bibliography{main}

\end{document}